\RequirePackage{lineno}
\documentclass[10pt]{IEEEtran}
\usepackage{epsfig,amsmath,amsfonts,cite}
\usepackage{threeparttable}
\usepackage{lipsum}
\usepackage{color}
\usepackage{graphicx}
\usepackage{lipsum}
\usepackage{algorithm,algorithmic}

\usepackage{booktabs}
\newcommand{\thickhline}{\noalign{\hrule height 1.0pt}}
\usepackage{multirow}%,multicol

\newcommand{\ten}[1]{\mathbfcal{#1}}   %mathcal
\newcommand{\mat}[1]{\mathbf{#1}}

\DeclareMathAlphabet\mathbfcal{OMS}{cmsy}{b}{n}

\newcommand{\parm}{{\xi}}
\newcommand{\vecpar}{\boldsymbol{\parm}}

\newcommand{\parNum}{d}

\newcommand{\out}{y}

\newcommand{\multiGPC}{\Psi }

\newcommand{\polyInd}{\alpha}
\newcommand{\basisInd}{\boldsymbol{\polyInd}}

\newcommand{\pcOrder}{p}

\newcommand{\yPC}{\sum\limits_{|\basisInd|=0}^{\pcOrder} {c_{\basisInd}  \multiGPC_{\basisInd}  (\vecpar)} }

\newcommand{\uvec}{\mathbf{u} }

\ifCLASSINFOpdf
 
\else
\fi

\begin{document}
%\title{Uncertainty Quantification for the Periodic Steady States of Forced and Autonomous Circuits via Stochastic Testing and Shooting Newton}
%\title{Generating Positive Approximation for Circuit Performance Density Using Rational Function Fitting and Semi-Definite Programming}
\title{A Big-Data Approach to Handle Many Process Variations: Tensor Recovery and Applications}
%\title{Uncertainty Quantification of Periodic Steady States for Analog, RF and Power Electronic Circuits}

\author{Zheng Zhang, Tsui-Wei Weng, and~Luca~Daniel
\thanks{Some preliminary results of this work have been reported in~\cite{zhang2016big}. This work was funded by the NSF NEEDS Program and by the AIM Photonics Program.}     
\thanks{Z. Zhang, T.-W. Weng and L. Daniel are with the Research Laboratory of Electronics, Massachusetts Institute of Technology (MIT), Cambridge, MA 02139, USA (e-mail: z\_zhang@mit.edu, twweng@mit.edu, luca@mit.edu).}
%\thanks{Copyright (c) 2015 IEEE. Personal use of this material is permitted. However, permission to use this material for any other purposes must be obtained from the IEEE by sending an email to pubs-permissions@ieee.org.}
}

\markboth{\MakeLowercase{Accepted by} IEEE TRANSACTIONS ON COMPONENT, PACKAGING AND MANUFACTURING TECHNOLOGY, ~Vol. ~XX, No.~XX,~XX~2016}{ZHANG \MakeLowercase{\textit{et al.}}: A Big Data Approach to Handle Process Variation}


\IEEEspecialpapernotice{(Invited Paper)}

\maketitle

\begin{abstract}
%\boldmath
Fabrication process variations are a major source of yield degradation in the nano-scale design of integrated circuits (IC), microelectromechanical systems (MEMS) and photonic circuits. Stochastic spectral methods are a promising technique to quantify the uncertainties caused by process variations. Despite their superior efficiency over Monte Carlo for many design cases, these algorithms suffer from the curse of dimensionality; i.e., their computational cost grows very fast as the number of random parameters increases. In order to solve this challenging problem, this paper presents a high-dimensional uncertainty quantification algorithm from a big-data perspective. Specifically, we show that the huge number of (e.g., $1.5 \times 10^{27}$) simulation samples in standard stochastic collocation can be reduced to a very small one (e.g., $500$) by exploiting some hidden structures of a high-dimensional data array. This idea is formulated as a tensor recovery problem with sparse and low-rank constraints; and it is solved with an alternating minimization approach. Numerical results show that our approach can simulate efficiently some ICs, as well as MEMS and photonic problems with over 50 independent random parameters, whereas the traditional algorithm can only handle several random parameters.
\end{abstract}


% \fi

\IEEEpeerreviewmaketitle

\section{Introduction}
\IEEEPARstart{F}{abrication} process variations (surface roughness of interconnects and nano-photonic devices, random doping effects of transistors) have become a critical issue in nano-scale design, because they can significantly influence chip performance and decrease product yield~\cite{variation2008}. In order to estimate and control the uncertainties in a design flow, efficient stochastic modeling and simulation algorithms should be developed and implemented in electronic design automation (EDA) software. For several decades, Monte Carlo techniques~\cite{MCintro, SingheeR09} have been the mainstream stochastic simulators in commercial tools due to their ease of implementation. Nevertheless, they have a slow convergence rate, and thus generally require a large number of repeated simulations. In recent years, the emerging stochastic spectral methods~\cite{sfem, col:2005} have been exploited in the EDA community, and they prove efficient for many design cases including integrated circuits (ICs)~\cite{manfredi:tcas2014, Stievano:2011_1, zzhang:tcad2013, Strunz:2008, zzhang:tcas2_2013,Pulch:2011_1, Rufuie2014, yucel2015me, ahadi2015sparse, ahadi2016hyperbolic,manfredi2015}, microelectromechanical systems (MEMS)~\cite{zzhang_cicc2014, zzhang:huq_tcad} and photonic circuits~\cite{twweng:optsEx, zubac2015efficient}.

The key idea of stochastic spectral methods is to approximate a stochastic solution (e.g., the uncertain voltage or power dissipation of a circuit) as the linear combination of some specialized basis functions such as generalized polynomial chaos~\cite{gPC2002}. Two main class of simulators have been implemented to obtain the coefficients of each basis functions. In an intrusive (i.e., non-sampling) simulator such as stochastic Galerkin~\cite{sfem} and stochastic testing~\cite{zzhang:tcad2013}, a new deterministic equation is constructed such that the unknown coefficients can be computed directly by a single simulation. Generally, stochastic testing~\cite{zzhang:tcad2013} is more efficient than stochastic Galerkin for many applications, since the resulting Jacobian matrix can be decoupled and the step sizes in transient analysis can be selected adaptively. %\textcolor{red}{The sampling-based variant of stochastic testing  is easier to implement, but it is less efficient for transient analysis due to the worse time step control~\cite{manfredi2015}}. 
In a sampling-based simulator such as stochastic collocation~\cite{col:2005}, a few solution samples are first computed by repeated simulations, then some post-processing techniques are used to reconstruct the unknown coefficients. The methods in~\cite{ahadi2015sparse, ahadi2016hyperbolic} reduce the complexity by selecting critical samples or critical basis functions. When the number of random parameters is small, these solvers can provide highly accurate solutions with significantly (e.g., $100\times$ to $1000\times$) higher efficiency than Monte Carlo. Unfortunately, stochastic spectral methods suffer from the curse of dimensionality, i.e. their computational cost grows very fast as the number of random parameters increases. 

\textbf{Related Work.} In order to solve high-dimensional problems, several advanced uncertainty quantification algorithms have been reported. Below are some representative high-dimensional solvers for IC and MEMS applications:
\begin{itemize}
	\item \textbf{Sparse Techniques.} In a high-dimensional polynomial chaos expansion, very often the coefficients of most basis functions are close to zero. In~\cite{zzhang_cicc2014}, this property was exploited for analog IC applications by using adaptive analysis of variance (ANOVA)~\cite{anchor_ANOVA_xiu:2012,anchor_ANOVA_xma:2010,HDMR:1999}. In~\cite{xli2010}, compressed sensing~\cite{Donoho:2006} was employed to minimize the $\ell _1$-norm of the coefficient vector. 
	\item \textbf{Matrix Low-Rank Approach.} In the intrusive solver reported in~\cite{Tarek_DAC:10}, all coefficient vectors of a stochastic solution were assembled as a matrix. The resulting matrix was found to have a low rank, and its most dominant factors were computed iteratively by nonlinear optimization. 
	\item \textbf{Model Order Reduction.} In~\cite{MoselhyD10}, an efficient reduced model was used to obtain most solution samples within a sampling-based solver. The reduced model is constructed by refinements. When a parameter value is detected for which the reduced model is inaccurate, the original large-scale equation is solved to update the model on-the-fly. 
	\item \textbf{Hierarchical Approach.} Using generalized polynomial-chaos expansions to describe devices and subsystems, the tensor-train hierarchical uncertainty quantification framework in~\cite{zzhang:huq_tcad} was able to handle complex systems with many uncertainties. The basic idea is to treat the stochastic output of each device/subsystem as a new random input. As a result, the system-level uncertainty quantification has only a small number of random parameters when new basis functions are used.
\end{itemize}

\textbf{Contributions.} This paper presents a sampling-based high-dimensional stochastic solver from a big-data perspective. The standard stochastic collocation approach was well known for its curse of dimensionality, and it was only applicable to problems with a few random parameters. In this paper, we represent the huge number of required solution samples as a tensor, which is a high-dimensional generalization of a matrix or a vector~\cite{tensor:suvey}\footnote{Tensor is an efficient tool to reduce the computational and memory cost of many problems (e.g., deep learning and data mining) in big-data analysis.}. In order to overcome the curse of dimensionality in stochastic collocation, we suggest a tensor recovery approach: we use a small number of simulation samples to estimate the whole tensor. This idea is implemented by exploiting the hidden low-rank property of a tensor and the sparsity of a generalized polynomial-chaos expansion. Numerical methods are developed to solve the proposed tensor recovery problem. We also apply this framework to simulate some IC, MEMS and photonic design cases with lots of process variations, and compare it with standard sampling-based stochastic spectral methods.

\textbf{Paper organization.} This paper is organized as follows. Section II briefly reviews stochastic collocation and tensor computation. Section III describes our tensor recovery model to reduce the computational cost of high-dimensional stochastic collocation. Numerical techniques are described in Section IV to solve the resulting optimization problem. Section V explains how to obtain a generalized polynomial-chaos expansion from the obtained tensor factors. The simulation results of some high-dimensional IC, MEMS and photonic circuit examples are reported in Section VI. Finally, Section VII concludes this paper and points out some future work.

\section{Preliminaries}
\subsection{Uncertainty Quantification using Stochastic Collocation}
\label{subsec:uq}
Let the vector $\vecpar=[\parm_1, \ldots, \parm_{\parNum}] \in \mathbb{R}^{\parNum}$ denote a set of mutually independent random parameters that describe process variations (e.g., deviation of transistor threshold voltage, thickness of a dielectric layer in MEMS fabrication). We intend to estimate the uncertainty of an output of interest $\out (\vecpar)$. This parameter-dependent output of interest can describe, for instance, the power consumption of an analog circuit, or the frequency of a MEMS resonator.

\textbf{Generalized Polynomial Chaos.} Assuming that $\out$ smoothly depends on $\vecpar$ and that $\out$ has a bounded variance\footnote{In this paper, we assume that $y$ is a scalar.}, we apply a truncated generalized polynomial-chaos expansion~\cite{gPC2002} to approximate the stochastic solution
\begin{equation}
\label{eq:ygpc}
\out (\vecpar) \approx \yPC, \; {\rm with}\; \mathbb{E}\left[{\multiGPC}_{\basisInd} \left( \vecpar \right) \multiGPC_{\boldsymbol{\beta }}\left( \vecpar \right)\right ]=\delta_{\basisInd, \boldsymbol{\beta }}.
\end{equation}
Here the operator $\mathbb{E}$ denotes expectation, $\delta$ denotes a Delta function, the basis functions $\{{\multiGPC}_{\basisInd} \left(\vecpar\right)\}$ are orthonormal polynomials, $\basisInd=[\alpha_1,\ldots, \alpha_{\parNum}] \in \mathbb{N}^{\parNum}$ is a vector indicating the highest polynomial order of each parameter in the corresponding basis. The total polynomial order $|\basisInd|=|\alpha_1|+\ldots +|\alpha_d|$ is bounded by $p$, and thus the total number of basis functions is $K=(p+d)!/(p!d!)$. Since $\vecpar$ are mutually independent, for  each parameter $\xi_k$ one can first construct a set of univariate orthonormal polynomials $ \phi_{k,\alpha_k} ( {\xi_k } ) $ with $\alpha_k=0,\ldots,p$. Then the multivariate polynomial basis function with index $\basisInd$ is 
\begin{equation}
\label{eq:mvgPC}
\multiGPC_{\basisInd}(\vecpar)=\prod\limits_{k=1}^d {\phi_{k,\alpha_k} ( {\xi_k } )}.
\end{equation}
The univariate polynomial functions can be obtained by the three-term recurrence relation in~\cite{Walter:1982}, and the main steps are summarized in Appendix~\ref{subsec:uni_gPC}.

\textbf{Stochastic Collocation.} Since all basis functions in \eqref{eq:ygpc} are orthonormal to each other, the coefficient $c_{\basisInd}$ can be obtained by a projection framework:
\begin{equation}
\label{eq:yproject}
c_{\basisInd}= %\mathbb{E}\left[ \out (\vecpar)  {\multiGPC}_{\basisInd}\right]=
\int\limits_{\mathbb{R}^d} {\out (\vecpar)  {\multiGPC}_{\basisInd} (\vecpar)\rho({\vecpar}) d\vecpar}, \;{\rm with}\; \rho(\vecpar)=\prod\limits_{k=1}^d {\rho _k(\xi _k)}.
\end{equation}
Note that $\rho(\vecpar)$ is the joint probability density function of vector $\vecpar$; $\rho_k(\xi_k)$ the marginal density of $\xi_k$. The above integral needs to be evaluated with a proper numerical technique. Popular integration techniques include randomized approaches such as Monte Carlo~\cite{MCintro}, and deterministic approaches like tensor product and sparse grid~\cite{Gerstner:1998}. Monte Carlo is feasible for extremely high-dimensional problems, but its numerical accuracy is low. Deterministic approaches can generate very accurate results by using a low-order quadrature rule, but they are only feasible for problems with a small or medium number of random parameters due to the curse of dimensionality.  This paper considers the tensor-product implementation, which was regarded as much less efficient than sparse grid techniques in almost all previous publications. 
\begin{figure}[t]
	\centering
		\includegraphics[width=3.3in]{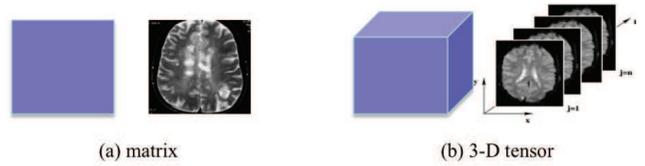} 
\caption{(a) a 2-D data array (e.g., a medical image) is a matrix, (b) a 3-D data array (e.g., multiple slices of images) is a tensor.}
	\label{fig:tensor}
\end{figure} 
\begin{figure*}[t]
	\centering
		\includegraphics[width=3.7in]{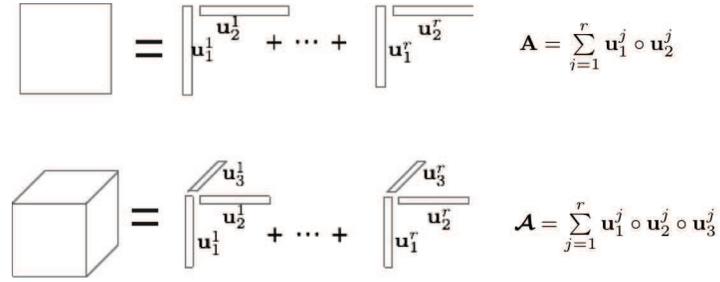} 
\caption{Low-rank factorization of a matrix (top), and the canonical decomposition of a third-order tensor (bottom).}
	\label{fig:tensor_fac}
\end{figure*}

We briefly introduce the idea of tensor-product numerical integration. Let $\{(\xi_k^{i_k}, w_k^{i_k})\}_{i_k=1}^n$ be $n$ pairs of 1-D quadrature points (or samples) and weights for parameter $\xi_k$. Such quadrature points and weights can be obtained by various numerical techniques, which can be found for instance in~\cite{Davis:07}. In this paper, we use the Gauss quadrature rule~\cite{Golub:1969} to generate such 1-D samples and weights, as summarized in Appendix~\ref{app:gauss_quad}. A Gauss quadrature rule with $n$ samples can generate exact results when the univariate integrand is a polynomial function of $\xi_k$ and when the highest polynomial degree is not higher than $2n-1$. By tensorizing all 1-D quadrature points/weights, the $d$-dimensional integral in (\ref{eq:yproject}) can be evaluated as
\begin{equation}
\label{eq:yTP}
c_{\basisInd}=\sum\limits_{1\leq i_1,\ldots, i_d\leq n} {\out (\vecpar_{i_1\ldots i_d})  {\multiGPC}_{\basisInd}  (\vecpar_{i_1\ldots i_d})  w_{i_1\ldots i_d}  }.
\end{equation}
Here $\vecpar_{i_1\ldots i_d}=[\xi_1^{i_1}, \ldots, \xi_d^{i_d}]$ and $w_{i_1\ldots i_d}=w_1^{i_1}\ldots w_d^{i_d}$ are the resulting multi-dimensional quadrature samples and weights, respectively. Obtaining each solution sample $\out (\vecpar_{i_1\ldots i_d})$ may require a time-consuming numerical simulation. For instance, a periodic steady-state solver may be called to compute the frequency of an oscillator. In device modeling, a large-scale solver must be called to solve a complex partial differential equation or integral equation for each quadrature sample. The numerical implementation \eqref{eq:yTP} requires $n^d$ times of such expensive device or circuit simulations. 

%The implementation in (\ref{eq:yTP}) is prohibitively expensive for high-dimensional case, because it needs evaluating $\out(\vecpar)$ at $q^d$ parameter samples. In practice, evaluating $y(\vecpar)$ at each sample can be time-consuming. For instance, a complicated periodic steady-state simulation must be run in RF circuit analysis~\cite{zzhang:tcas2_2013}; a large-scale PDE or integral equation must be solved in device modeling~\cite{MoselhyD10}. Therefore, it is always desired to utilize as few simulation samples as possible.

%We first summarize the procedures of constructing gPC basis functions.

\subsection{Tensor and Tensor Decomposition}
\subsubsection{Tensor} Tensor is a high-dimensional generalization of matrix. A matrix $\mat{X} \in \mathbb{R}^{n_1\times n_2}$ is a $2$nd-order tensor, and its element indexed by $\mat{i}=(i_1, i_2)$ can be denoted as $x_{i_1i_2}$ or $\mat{X}(\mat{i})$. For a general $d$th-order tensor $\ten{X} \in \mathbb{R}^{n_1\times \ldots n_d}$, its element indexed by $\mat{i}=(i_1, \ldots, i_d)$ can be denoted as $x_{i_1\ldots i_d}$ or $\ten{X}(\mat{i})$. Here the integer $k \in [1,d]$ is the index for a \textbf{mode} of $\ten{X}$. Fig.~\ref{fig:tensor} shows a matrix and  a third-order tensor. 

Given any two tensors $\ten{X}$ and $\ten{Y}$ of the same size, their inner product is defined as
\begin{equation}
\langle \ten{X}, \ten{Y} \rangle :=\sum\limits_{i_1\ldots i_d} {x_{i_1\ldots i_d} y_{i_1\ldots i_d}}. \nonumber
\end{equation}
The Frobenius norm of tensor $\ten{X}$ is further defined as $|| \ten{X}||_F :=\sqrt{\langle \ten{X}, \ten{X} \rangle}$.

\subsubsection{Tensor Decomposition} A tensor $\ten{X} $ is rank-1 if it can be written as the outer product of some vectors:
\begin{equation}
\label{eq:tensor_rank1}
\ten{X}=\uvec_1 \circ \cdots \circ \uvec_{\parNum}\; \Leftrightarrow\; x_{i_1\ldots i_{\parNum}}=\uvec_1(i_1) \cdots  \uvec_{\parNum}(i_{\parNum})
\end{equation}
where $\mat{u}_k(i_k)$ denotes the $i_k$-th element of vector $\mat{u}_k \in  \mathbb{R}^{n_k}$. Similar to matrices, a low-rank tensor can be written as a canonical decomposition~\cite{candecomp}, which expresses $\ten{X}$ as the sum of some rank-1 tensors:
\begin{equation}
\label{eq:tensor_rank_r}
\ten{X}=\mathbb{T}_{\rm cp}\left( \mat{U}^{(1)}, \ldots, \mat{U}^{(d)} \right):=\sum\limits_{j=1}^r {\uvec_1^j \circ \cdots \circ \uvec_{\parNum}^j}.
\end{equation}
Here $\mat{U}^{(k)}=[\mat{u}_k^1, \ldots, \mat{u}_k^r]\in \mathbb{R}^{n_k \times r}$ is a matrix including all factors corresponding to mode $k$; operator $\mathbb{T}_{\rm cp}$ converts all matrix factors to a tensor represented by canonical decomposition; the minimum integer $r$ that ensures \eqref{eq:tensor_rank_r} to hold is called \textbf{tensor rank}. As a demonstration, Fig.~\ref{fig:tensor_fac} shows the low-rank factorizations of a matrix and third-order tensor, respectively. Tensor decomposition \eqref{eq:tensor_rank_r} can significantly reduce the cost of storing high-dimensional data arrays. For simplicity, let us assume $n_k=n$. Directly representing tensor $\ten{X}$ requires storing $n^d$ scalars, whereas only $ndr$ scalars need to be stored if the above low-rank factorization exists. 

Note that there are other kinds of tensor factorizations such as Tucker decomposition~\cite{tuckerreview} and tensor-train decomposition~\cite{ivanTT}. We only introduce canonical decomposition in this paper because we will use it to solve high-dimensional uncertainty quantification problems in the subsequent sections. Interested readers are referred to~\cite{tensor:suvey} for a detailed survey of tensor decompositions, as well as \cite{zhang:2016keynote} for a tutorial with applications in electronic design automation.

\section{Tensor Recovery Approach}
Formulation (\ref{eq:yTP}) was only applicable to problems with $5$ or $6$ random parameters due to the $n^d$ simulation samples. This section describes our tensor-recovery approach that can significantly reduce the computational cost of tensor-product stochastic collocation. With this framework, (\ref{eq:yTP}) can be more efficient than sparse-grid approaches and Monte Carlo simulation for many high-dimensional design cases.

%\subsection{Stochastic Collocation Using Tensor Recovery}

\subsection{Reformulating Stochastic Collocation with Tensors} 

We first define the following two tensors:
\begin{itemize}
	\item tensor $\ten{Y} \in \mathbb{R}^{n_1\times\ldots \times n_d}$, with $n_k$$=$$n$ and each element being $y_{i_1\ldots i_d}=y(\vecpar_{i_1\ldots i_d })$;
	\item tensor $ \ten{W}_{\basisInd} \in \mathbb{R}^{n_1\times\ldots \times n_d}$, with $n_k$$=$$n$ and its element indexed by $(i_1, \ldots i_d)$ being ${\multiGPC}_{\basisInd}  (\vecpar_{i_1\ldots i_d})  w_{i_1\ldots i_d}$.
\end{itemize}
Tensor $ \ten{W}_{\basisInd} $ only depends on the basis function in \eqref{eq:ygpc} and the multi-dimensional quadrature weights in \eqref{eq:yTP}. Furthermore, according to \eqref{eq:mvgPC}, it is straightforward to see that $ \ten{W}_{\basisInd} $ is a rank-1 tensor with the following canonical decomposition:
\begin{align}
\label{eq:tensorW}
&\ten{W}_{\basisInd}  =\mat{v}_1^{\alpha_1} \circ \cdots \circ \mat{v}_{\parNum}^{\alpha_d}, \nonumber \\
& {\rm with}\; \mat{v}_k^{\alpha_k}=[\phi_{k,\alpha_k} ( {\xi_k^{1} } )w_k^{1},\ldots, \phi_{k,\alpha_k} ( {\xi_k^{n} } )w_k^{n}]^T \in \mathbb{R}^{n\times 1}.
\end{align}
Note that $\xi_k^{i_k}$ and $w_k^{i_k}$ are the $i_k$th 1-D quadrature point and weight for parameter $\xi_k$, as described in Section~\ref{subsec:uq}.

With the above two tensors, Equation (\ref{eq:yTP}) can be written in the following compact form:
\begin{equation}
\label{eq:cTensor}
c_{\basisInd} =\langle \ten{Y}, \ten{W}_{\basisInd} \rangle.
\end{equation}
 \begin{figure*}[ht]
 \begin{align}
\label{tensor_lrsp}
 \min_{\mat{U}^{(1)}, \ldots, \mat{U}^{(d)} \in \mathbb{R}^{n\times r} } &\; f\left( \mat{U}^{(1)}, \ldots, \mat{U}^{(d)} \right) \nonumber \\
  & =\frac{1}{2} \left \| \mathbb{P}_{\Omega}\left(\mathbb{T}_{\rm cp} \left(\mat{U}^{(1)}, \ldots, \mat{U}^{(d)} \right)  -{\ten{Y}}\right)\right \|_F^2 +\lambda \sum\limits_{|\basisInd|=0}^p \left |\left\langle \mathbb{T}_{\rm cp} \left(\mat{U}^{(1)}, \ldots, \mat{U}^{(d)} \right), \ten{W}_{\basisInd}\right\rangle \right |.
\end{align}
\end{figure*}
Since $ \ten{W}_{\basisInd} $ is straightforward to obtain, the main computational cost is to compute $ \ten{Y}$. Once $ \ten{Y}$ is computed, $c_{\basisInd}$'s and thus the generalized polynomial-chaos expansion \eqref{eq:ygpc} can be obtained easily. Unfortunately, directly computing $\ten{Y}$ is impossible for high-dimensional cases, since it requires simulating a specific design case $n^d$ times.

\subsection{Tensor Recovery Problem (Ill-Posed)} 
We define two index sets:
\begin{itemize}
	\item Let ${\cal I}$ include all indices $(i_1,\ldots, i_d )$ for the elements of $\ten{Y}$. The number of elements in ${\cal I}$, denoted as  $|{\cal I}|$, is $n^d$;
	\item Let $\Omega$ be a small subset of ${\cal I}$, with $|{\Omega}|\ll|{\cal I}|$. For each index $(i_1,\ldots, i_d) \in \Omega$, the corresponding solution sample $y_{i_1\ldots i_d}$ is already obtained by a numerical simulator.
\end{itemize}
In order to reduce the computational cost, we aim at estimating the whole tensor $ \ten{Y}$ using the small number of available simulation data specified by $\Omega$. With the sampling set $\Omega$, a projection operator $\mathbb{P}$ is defined for $\ten{Y}$:
\begin{equation}
\label{tensor_project}
\ten{B}=\mathbb{P}_{\Omega}\left({\ten{Y}}\right) \; \Leftrightarrow\;  b_{i_1\ldots i_d} = \left\{ \begin{array}{l}
 y_{i_1\ldots i_d} ,\;{\rm{if}}\; (i_1,\ldots, i_d) \in  {\Omega} \\
 0 ,\;{\rm{otherwise}}.
 \end{array} \right.
 \end{equation}
We want to find a tensor $\ten{X}$ such that it matches $\ten{Y}$ for the elements specified by $\Omega$:
 \begin{equation}
\label{tensor_point_match}
\| \mathbb{P}_{\Omega}\left(\ten{X} -{\ten{Y}}\right)\|_F^2 =0.
\end{equation}
However, this problem is \textbf{ill-posed}, because any value can be assigned to $x_{i_1\ldots i_d}$ if $(i_1,\ldots, i_d) \notin \Omega$.

\subsection{Regularized Tensor Recovery Model} 

In order to make the tensor recovery problem well-posed, we add the following constraints based on heuristic observations and practical implementations.
\begin{itemize}
\item \textbf{Low-Rank Constraint.} Very often we observe that the high-dimensional solution data array $\ten{Y}$ has a low tensor rank. Therefore, we expect that its approximation $\ten{X}$ has a low-rank decomposition described in (\ref{eq:tensor_rank_r}).

\item \textbf{Sparse Constraint.} As shown in previous work of compressed sensing~\cite{xli2010} and ANOVA decomposition~\cite{zzhang_cicc2014}, most of the coefficients in a high-dimensional generalized polynomial-chaos expansion have very small magnitude. This implies that the $\ell_1$-norm of a vector collecting all coefficients $c_{\basisInd}$'s, which is computed as
\begin{equation}
\sum\limits_{|\basisInd|=0}^p{| c_{\basisInd} |} \approx \sum\limits_{|\basisInd|=0}^p{| \langle \ten{X}, \ten{W}_{\basisInd}\rangle |}, 
\end{equation}
should be very small.
\end{itemize}
  
\textbf{Finalized Tensor Recovery Model.} Combining the above low-rank and sparse constraints together, we suggest the finalized tensor recovery model  \eqref{tensor_lrsp} to compute $\ten{X}$ as an estimation of $\ten{Y}$. In this formulation, $\ten{X}$ is assumed to have a rank-$r$ decomposition, and we compute its matrix factors $\mat{U}^{(k)}$'s instead of the whole tensor $\ten{X}$. This treatment has a significant advantage: the number of unknown variables is reduced from $n^d$ to $dnr$, which is now a linear function of parameter dimensionality $d$.

\subsection{Cross Validation}
 An interesting question is: how accurate is $\ten{X}$ compared with the exact tensor $\ten{Y}$? Our tensor recovery formulation enforces consistency between $\ten{X}$ and $\ten{Y}$ at the indices specified by $\Omega$. It is desired that $\ten{X}$ also has a good predictive behavior -- $x_{i_1\ldots i_d}$ is also close to $x_{i_1\ldots i_d}$ for $(i_1,\ldots, i_d) \notin \Omega$. In order to measure the predictive property of our results, we define a heuristic prediction error
\begin{equation}
\epsilon_{\rm pr}=\sqrt{\frac{\sum \limits_{(i_1,\ldots, i_d)\in \Omega'}{\left(x_{i_1\ldots i_d} - y_{i_1\ldots i_d} \right)^2 w_{i_1\ldots i_d}}}  {\sum \limits_{(i_1,\ldots, i_d)\in \Omega'}{\left( y_{i_1\ldots i_d} \right)^2 w_{i_1\ldots i_d}}}}. \nonumber
\end{equation}
Here $\Omega' \subset {\cal I}$ is a small-size index set such that $\Omega' \cap \Omega=\emptyset$. The solution $\ten{X}$ is regarded as a good approximate to $\ten{Y}$ if $\epsilon_{\rm pr}$ is small; then \eqref{eq:ygpc} can be obtained accurately by using \eqref{eq:cTensor}, and the statistical behavior (e.g., probability density function) of $\out (\vecpar) $ can be well predicted. Estimating $\epsilon_{\rm pr}$ requires simulating the design problem at some extra quadrature samples. However, a small-size $\Omega' $ can provide a good heuristic estimation.

At present, we do not have a rigorous approach to find the optimal values of $\lambda$ and $r$. In practice, their values are chosen heuristically. Specifically, we increment $\lambda$ and $r$ until $\epsilon_{\rm pr}$ becomes small enough. Occasionally the optimization problem \eqref{tensor_lrsp} may be solved several times for different values of $\lambda$ and $r$. However, like other sampling-based stochastic solvers, the computational cost of post-processing [i.e., solving \eqref{tensor_lrsp}] is generally negligible compared with the cost of simulating solution samples indexed by $\Omega$.

\section{Solve Problem \eqref{tensor_lrsp}} 
\label{sec:opt_solver}

The optimization problem \eqref{tensor_lrsp} is solved iteratively in our implementation. Specifically, starting from a provided initial guess of the low-rank factors $\{\mat{U}^{(k)} \}_{k=1}^d$, alternating minimization is performed recursively using the result of the previous iteration as a new initial guess. Each iteration of alternating minimization consists of $d$ steps. At the $k$-th step, the $k$th-mode factor matrix $\mat{U}^{(k)}$ corresponding to parameter $\xi_k$ is updated by keeping all other factors fixed and by solving (\ref{tensor_lrsp}) as a convex optimization problem.

\subsection{Outer Loop: Alternating Minimization}
\textbf{Algorithm Flow.} We use an iterative algorithm to solve (\ref{tensor_lrsp}). Let $\mat{U}^{(k),l}$ be the mode$-k$ factors of $\ten{X}$ after $l$ iterations. Starting from an initial guess $\{\mat{U}^{(k),0}\}_{k=1}^d$, we perform the following iterations:
\begin{itemize}
	\item At iteration $l+1$, we use $\{\mat{U}^{(k),l}\}_{k=1}^d$ as an initial guess and obtain updated tensor factors $\{\mat{U}^{(k),l+1}\}_{k=1}^d$ by alternating minimization. 
	\item Each iteration consists of $d$ steps; at the $k$-th step, ${\mat{U}^{(k),l+1}}$ is obtained by solving
	\small
\begin{equation}
\label{eq:alm_k}
{\mat{U}^{(k),l+1}}=\arg \min_{\mat{X}} { f\left( \ldots, \mat{U}^{(k-1),l+1}, \mat{X}, \mat{U}^{(k+1),l},\ldots \right)}.
\end{equation} \normalsize
\end{itemize}
Since all factors except that of mode $k$ are fixed, (\ref{eq:alm_k}) becomes a convex optimization problem, and its global minimum can be computed by the solver in Section~\ref{subsec:admm}. The alternating minimization method ensures that the cost function decreases monotonically to a local minimal. The pseudo codes are summarized in Alg.~\ref{alg:alm}.

\begin{algorithm}[t]
\caption{Alternating Minimization for Solving \eqref{tensor_lrsp}.}
\label{alg:alm}
\begin{algorithmic}[1]
\STATE {Initialize:  $\mat{U}^{(k),0}\in \mathbb{R}^{n\times r}$ for $k=1,\ldots d$;}
\STATE {\textbf{for} $l=0,1,\ldots$}
  \STATE {\hspace{10pt} {\textbf{for} $k=1,\;\ldots, d$  \textbf{do}}}
   \STATE {\hspace{20pt} solve (\ref{eq:alm_k}) by Alg. \ref{alg:admm} to obtain $\mat{U}^{(k),l+1}$ };
   \STATE {\hspace{10pt} {\textbf{end for} } }
   \STATE {\hspace{10pt} \textbf{break} if converged;}
  \STATE {\textbf{end for} } 
\STATE {\textbf{return} $\mat{U}^{(k)}=\mat{U}^{(k),l+1}$ for $k=1,\ldots, d$. }
\end{algorithmic}
\end{algorithm}

\textbf{Convergence Criteria.} With matrices $\{ \mat{U}^{(k), l}\}_{k=1}^d$ obtained after $l$ iterations of the outer loops of Alg.~\ref{alg:alm}, we define
\begin{align}
f_l  & :=f\left( \mat{U}^{(1), l}, \ldots,   \mat{U}^{(d), l} \right),  \nonumber \\
\ten{X} _l & :=\mathbb{T}_{\rm cp}  \left( \mat{U}^{(1),l}, \ldots, \mat{U}^{(d),l} \right),  \nonumber\\
c_{\basisInd}^l&:=\left\langle \ten{X}_l, \mathbfcal{W}_{\basisInd} \right\rangle. \nonumber
\end{align} 
The first term is the updated cost function of \eqref{tensor_lrsp}; the second term is the updated tensor solution; the last term is the updated coefficient corresponding to basis function ${\multiGPC}_{\basisInd} (\vecpar)$ in \eqref{eq:ygpc}. Let $\mat{c}^l =[\ldots,  c_{\basisInd}^l, \ldots ]\in \mathbb{R}^K$ collect all coefficients in \eqref{eq:ygpc}, then we define the following quantities for error control:
 \begin{itemize}
 \item Relative update of the tensor factors:
 \begin{align} 
\epsilon_{l, \rm tensor} & =\sqrt{ \frac{{\sum \limits_{k=1}^d{\| \mat{U}^{(k),l} -\mat{U}^{(k),l-1} \|_F^2} }}{ { \sum \limits_{k=1}^d{\| \mat{U}^{(k),l-1}} \|_F^2}  }} .\nonumber
\end{align}

\item Relative update of $\mat{c}=[\ldots, c_{\basisInd},\ldots]$
 \begin{align} 
\epsilon_{l, \rm gPC}  = { \|  \mat{c}^l-  \mat{c}^{l-1} \|   }  /  {\| \mat{c}^{l-1}\|} .  \nonumber
\end{align} 

\item Relative update of the cost function:
 \begin{align} 
\epsilon_{l,\rm cost} &= { |  f_l-   f_{l-1} | }  /  {|f_{l-1}|}. \nonumber
\end{align} 

 \end{itemize}
The computed factors $ \mat{U}^{(1), l}, \ldots, \mat{U}^{(d), l}$ are regarded as a local minimal and thus Alg.~\ref{alg:alm} terminates if $\epsilon_{l, \rm tensor} $, $\epsilon_{l, \rm gPC} $ and $\epsilon_{l, \rm cost} $ are all small enough.

\begin{algorithm}[t]
\caption{ADMM for Solving (\ref{eq:alm_k}).}
\label{alg:admm}
\begin{algorithmic}[1]
\STATE {Initialize:  form $\mat{A,F}$ and $\mat{b}$ according to Appendix~\ref{app:glasso}, specify initial guess $\mat{x}^0$, $\mat{u}^0$ and $\mat{z}^0$;}
\STATE {\textbf{for} $j=0,1,\ldots$ \textbf{do}}
   \STATE {\hspace{5pt} compute $\mat{x}^{j+1}$, $\mat{z}^{j+1}$ and $\mat{u}^{j+1}$ according to (\ref{eq:dmm_update})};
      \STATE {\hspace{5pt} \textbf{break} if $\|\mat{Fx}^{j+1}-\mat{z}^{j+1}\|<\epsilon_1$ \& $\|\mat{F}^T(\mat{z}^{j+1}-\mat{z}^{j})\|<\epsilon_2$;}
  \STATE {\textbf{end for} } 
\STATE {\textbf{return} $\mat{U}^{(k),l+1}={\rm reshape}(\mat{x}^{j+1}, [n,r])$ . }
\end{algorithmic}
\end{algorithm}

\subsection{Inner Loop: Subroutine for Solving (\ref{eq:alm_k})}
\label{subsec:admm}

Following the procedures in Appendix~\ref{app:glasso}, we rewrite Problem (\ref{eq:alm_k}) as a generalized LASSO problem:
\begin{equation}
\label{eq:glasso}
\mat{vec}\left(\mat{U}^{(k),l+1} \right) =\arg \min_{\mat{x}} { \frac{1}{2} \left\| \mat{A}\mat{x}-\mat{b} \right\|_2^2 +\lambda | \mat{Fx}| }
\end{equation}
where $\mat{A}\in \mathbb{R}^{|\Omega| \times nr}$, $\mat{F}\in \mathbb{R}^{K \times nr}$ and $\mat{b}\in \mathbb{R}^{ |\Omega| \times 1}$, and $\mat{x}=\mat{vec}(\mat{X})\in \mathbb{R}^{nr \times 1}$ is the vectorization of $\mat{X}$ [i.e., the $(jn-n+i)$th element of $\mat{x}$ is $ \mat{X} (i,j)$ for any integer $1\leq i\leq n$ and $1\leq j \leq r$]. Note that $|\Omega|$ is the number of available simulations samples in tensor recovery, and $K=(p+d)!/(p!d!)$ is the total number of basis functions in \eqref{eq:ygpc}. 
\begin{figure*}[t]
	\centering
		\includegraphics[width=5.0in]{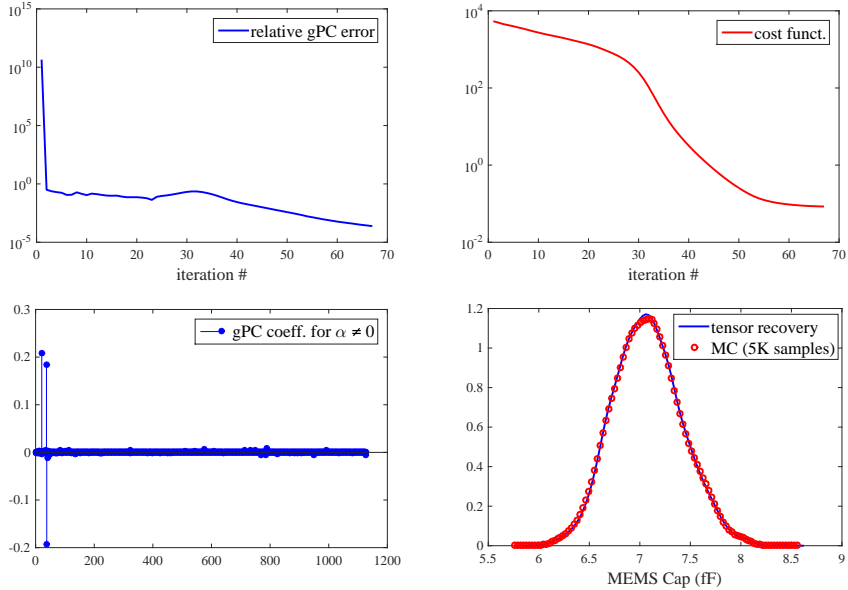} 
\caption{Numerical results of the MEMS capacitor, with $r=3$ and $\lambda=0.01$. Top left: relative error of the generalized polynomial-chaos coefficients in iterations; top right: decrease of the cost function in (\ref{tensor_lrsp}); bottom left: sparsity of the obtained generalized polynomial-chaos expansion; bottom right: obtained probability density function compared with that from Monte Carlo. }
	\label{fig:mems_results}
\end{figure*}

We solve (\ref{eq:glasso}) by the alternating direction method of multipliers (ADMM)~\cite{Boyd:ADMM2010}. Problem (\ref{eq:glasso}) can be rewritten as
\begin{equation}
%\label{eq:glasso_admm}
\min_{\mat{x,z}} { \frac{1}{2} \left\| \mat{A}\mat{x}-\mat{b} \right\|_2^2 +\lambda | \mat{z}| } \;\;\;\;\; {\rm s. } {\rm t.} \; \mat{Fx-z=0} .\nonumber 
\end{equation}
By introducing an auxiliary variable $\mat{u} $ and starting with initial guesses $\mat{x}^0$, $\mat{u}^0=\mat{z}^0=\mat{Fx}^0$, the following iterations are performed to update $\mat{x}$ and $\mat{z}$:
\begin{align}
\label{eq:dmm_update}
\mat{x}^{j+1} &=\left( \mat{A}^T\mat{A}+s\mat{F}^T \mat{F}\right)^{-1} (   \mat{A}^T\mat{b}+   s\mat{F}^T(\mat{z}^j-\mat{u}^j) )\nonumber \\
\mat  {z}^{j+1} &={\rm shrink}_{\lambda/s}(\mat{Fx}^{j+1}+\mat{z}^{j}+\mat{u}^j)\\
\mat  {u}^{j+1} &=\mat{u}^j+\mat{Fx}^{j+1}-\mat{z}^{j+1}.\nonumber
\end{align}
Here $s>0$ is an augmented Lagrangian parameter, and the soft thresholding operator is defined as
\begin{equation}
{\rm shrink}_{\lambda/s}(a)= \left\{ \begin{array}{l}
 a-\lambda/s,\; {\rm{if}}\; a> \lambda/s \\ 
 0 ,\;\;\;\;\; \;\;\;\;\;\;{\rm{if}}\;|a|< \lambda/s \\ 
 a+\lambda/s, \; {\rm{if}}\; a<- \lambda/s. 
 \end{array} \right.\nonumber
\end{equation}

The pseudo codes for solving (\ref{eq:alm_k}) are given in Alg.~\ref{alg:admm}.

\subsection{Remarks}
\label{subsec:limitation}
The cost function of \eqref{tensor_lrsp} is non-convex, therefore it is non-trivial to obtain a globally optimal solution with theoretical guarantees. Theoretically speaking, the numerical solution of a non-convex optimization problem depends on the given initial guess. Although researchers and engineers are very often satisfied with a local minimal, the obtained result may not be good enough for certain cases. In our examples, we find that using random initial guess works well for most cases. However, novel numerical solvers are still highly desired to compute the globally optimal solution of \eqref{tensor_lrsp} with theoretical guarantees.

\section{Generating Stochastic Model (\ref{eq:ygpc})}
Assuming that the low-rank factors $\mat{U}^{(1)}, \ldots, \mat{U^{(d)}}$ of $\ten{X}$ have been computed, we are ready to compute the coefficient $c_{\basisInd}$ for each basis function in \eqref{eq:ygpc}. Specifically, replacing $\ten{Y}$ with $\ten{X}$ in \eqref{eq:cTensor}, and exploiting the rank-1 property of $\ten{W}_{\basisInd}$ in \eqref{eq:tensorW}, we can easily compute $c_{\basisInd}$ by
\begin{align}
c_{\basisInd} & \approx \langle \ten{X}, \ten{W}_{\basisInd} \rangle  = \sum\limits_{j=1}^r{ \left( \prod\limits_{k=1}^d { \langle \mat{u}_k^j, \mat{v}_k^{\alpha_k}  \rangle  } \right) } \nonumber
\end{align} 
where $\mat{v}_k^{\alpha_k}$ is a low-rank factor of $\ten{W}_{\basisInd}$ in \eqref{eq:tensorW}. The above expression can be computed by efficient vector inner products. 

Once the generalized polynomial-chaos expansion \eqref{eq:ygpc} is obtained, various statistical information of the performance metric $y(\vecpar)$ can be obtained. For instance, the expectation and standard deviation of $y(\vecpar)$ can be obtained analytically; the density function of $y(\vecpar)$ can be obtained by sampling \eqref{eq:ygpc} or by using the maximum-entropy algorithm~\cite{zhang2014maximum}.  
\begin{figure}[t]
	\centering
		\includegraphics[width=2.4in]{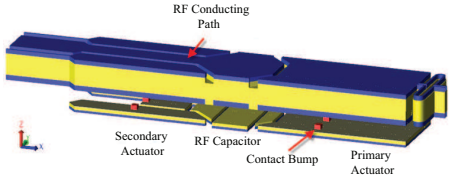} 
\caption{Schematic of a RF MEMS capacitor~\cite{zzhang:JMEMS2014}.}
	\label{fig:memsCap}
\end{figure}

\section{Numerical results}
In order to verify the effectiveness of our tensor-recovery uncertainty quantification framework, we show the simulation results of three examples ranging from integrated circuits, MEMS and photonic circuits. Since our focus is to solve high-dimensional problems, we simply assume that all process variations are mutually independent to each other, although they are likely to be correlated in practice. All codes are implemented in MATLAB and run on a Macbook with 2.5-GHz CPU and 16-G memory.
\begin{figure*}[t]
	\centering
		\includegraphics[width=5.0in]{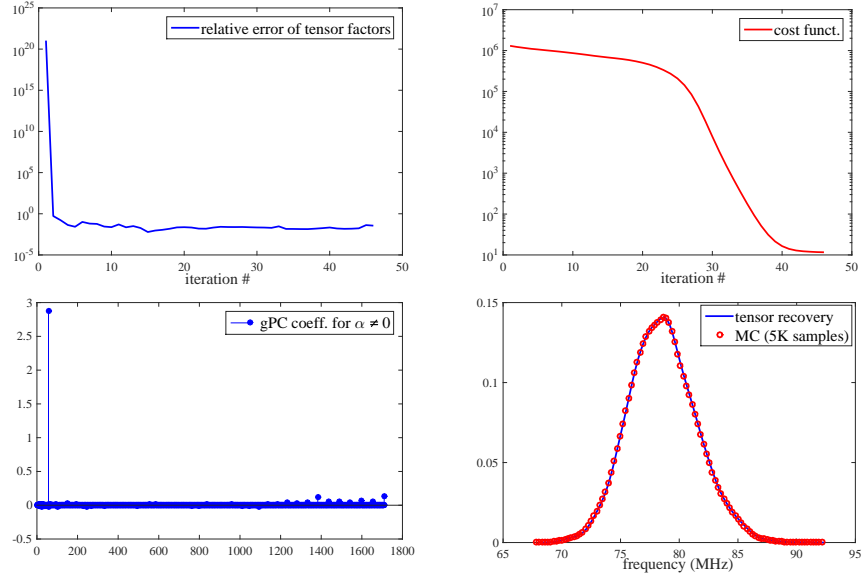} 
\caption{Numerical results of the ring oscillator, with $r=3$ and $\lambda=0.1$. Top left: relative error of the tensor factors for each iteration; top right: decrease of the cost function in (\ref{tensor_lrsp}); bottom left: sparsity of the obtained generalized polynomial-chaos expansion; bottom right: obtained density function compared with that from Monte Carlo using $5000$ samples. }
	\label{fig:ring_results}
\end{figure*}

\subsection{MEMS Example (with 46 Random Parameters)}
\begin{table} [t]
\caption{Comparison of simulation cost for the MEMS capacitor.} 
\centering 
\begin{tabular}{c c c c} 
\hline
method & tensor product & sparse grid & proposed \\  \thickhline 
simulation samples & $ 8.9\times 10^{21}$ & $4512$ & $300$ \\
\hline %inserts single line 
\end{tabular} 
\label{table:mems_cost}% is used to refer this table in the text  
\end{table} 
We first consider the MEMS device in Fig.~\ref{fig:memsCap}, which was described with details in~\cite{zzhang:JMEMS2014}. This example has $46$ random parameters describing the material and geometric uncertainties in CMOS fabrication. The capacitance of this device depends on both bias voltage and process parameters. We assume that a fixed DC voltage is applied to this device, such that we can approximate the capacitance as a $2$nd-order generalized polynomial-chaos expansion of $46$ random parameters. Assume that we use $3$ Gauss-quadrature points for each parameter. Consequently, as shown in Table~\ref{table:mems_cost}, a tensor-product integration requires $3^{46}\approx 8.9\times 10^{21}$ simulation samples, and the Smolyak sparse-grid technique still requires $4512$ simulation samples. 

We simulate this device using only $300$ quadrature samples randomly selected from the tensor-product integration rules, then our tensor recovery method estimates the whole tensor $\ten{Y}$ [which contains all $3^{46}$ samples for the output $\out (\vecpar)$]. The relative approximation error for the whole tensor is about $0.1\%$ (measured by cross validation).  As shown in Fig.~\ref{fig:mems_results}, our optimization algorithm converges with less than $70$ iterations, and the generalized polynomial-chaos coefficients are obtained with a small relative error (below $10^{-4}$); the obtained model is very sparse, and the obtained density function of the MEMS capacitor is almost identical with that from Monte Carlo. Note that the number of repeated simulations in our algorithm is only about $1/4$ of the total number of basis functions. 

\begin{figure}[t]
	\centering
		\includegraphics[width=2.3in]{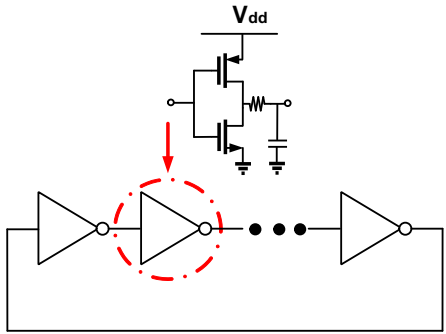} 
\caption{Schematic of a CMOS ring oscillator. }
	\label{fig:ring}
\end{figure}

\subsection{Multi-Stage CMOS Ring Oscillator (with 57 Parameters)}
We continue to consider the CMOS ring oscillator in Fig.~\ref{fig:ring}. This circuit has $7$ stages of CMOS inverters; $57$ random parameters are used to describe the variations of threshold voltages, gate-oxide thickness, and effective gate length/width. We intend to obtain a $2$nd-order polynomial-chaos expansion for its oscillation frequency by calling a periodic steady-state simulator repeatedly. The required number of simulations for different algorithms are listed in Table~\ref{table:ring_cost}, which clearly shows the superior efficiency of our approach for this example.

\begin{table} [t]
\caption{Comparison of simulation cost for the ring oscillator.} 
\centering 
\begin{tabular}{c c c c} 
\hline
method & tensor product & sparse grid & proposed \\  \thickhline % inserts single horizontal line 
simulation samples & $ 1.6\times 10^{27}$ & $6844$ & $500$ \\ \hline %inserts single line 
\end{tabular} 
\label{table:ring_cost}% is used to refer this table in the text  
\end{table}

We simulate this circuit using only $500$ samples randomly selected from the $3^{57} \approx 1.6\times 10^{27}$ tensor-product integration samples, then our algorithm estimates the whole tensor $\ten{Y}$ with a $1\%$ relative error.  As shown in Fig.~\ref{fig:ring_results}, our optimization algorithm converges after $46$ iterations, and the tensor factors are obtained with less than $1\%$ relative errors; the obtained model is very sparse, and the obtained density function of the oscillator frequency is almost identical with that from Monte Carlo. Note that the number of our simulations (i.e., $500$) is much smaller than the total number of basis functions (i.e., $1711$) in the generalized polynomial-chaos expansion.

\begin{figure*}[t]
	\centering
		\includegraphics[width=5.0in]{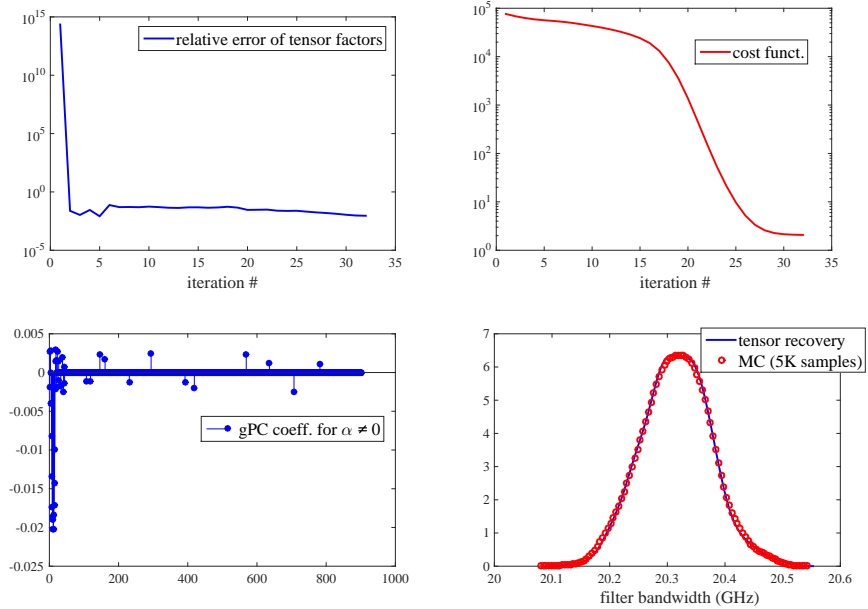} 
\caption{Numerical results of the photonic bandpass filter, with $r=3$, $\lambda=0.1$. Top left: relative error of the tensor factors for each iteration; top right: decrease of the cost function in (\ref{tensor_lrsp}); bottom left: sparsity of the obtained generalized polynomial-chaos expansion; bottom right: obtained density function of the filter bandwidth compared with that from Monte Carlo using $5000$ samples. }
	\label{fig:optics_results}
\end{figure*}

\subsection{Photonic Bandpass Filter (with 41 Parameters)}

\begin{figure}[t]
	\centering
		\includegraphics[width=2.3in]{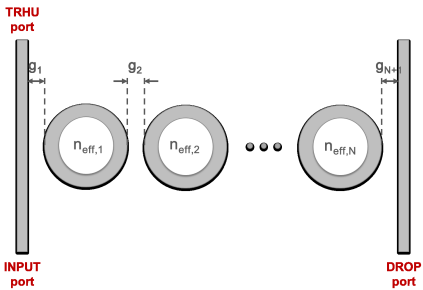} 
\caption{Schematic of a photonic bandpass filter, with $N=20$. }
	\label{fig:photonics}
\end{figure}

Finally we consider the photonic bandpass filter in Fig~\ref{fig:photonics}. This Chebyshev-type filter has $20$ ring resonators, and was originally designed to have a 3-dB bandwidth of 20 GHz, a 26-dB minimum return loss, a 400-GHz free spectral range, and a 1.55-$\mu$m operation wavelength. A total of $41$ random parameters are used to describe the variations of the effective phase index ($n_{\rm eff}$) of each ring, as well as the gap ($g$) between adjoint rings and between the first/last ring and the bus waveguides. These parameters are assumed to be independent Gaussian variables, with $n_{{\rm eff},i} = 2.2315585 + {\cal N}(0,5\times 10^{-6})$, and $g_i = 0.3 + {\cal N}(0,10^{-3}) \mu$m. We intend to obtain a $2$nd-order polynomial-chaos expansion for the 3-dB bandwidth at the DROP port of this filter. The required number of simulations for different algorithms are listed in Table~\ref{table:optics_cost}. Similar to the results of previous two examples, our tensor-recovery approach is significantly more efficient than the standard tensor-product stochastic collocation and the sparse-grid implementation.

We simulate this photonic circuit using only $500$ samples randomly selected from the $3^{41} \approx 3.6\times 10^{19}$ tensor-product integration samples, then our algorithm estimates the whole tensor $\ten{Y}$ with a $0.1\%$ relative error.  As shown in Fig.~\ref{fig:optics_results}, our optimization algorithm converges after $32$ iterations, and the tensor factors are obtained with less than $1\%$ relative errors; the obtained model is also sparse, and the obtained density function of the bandwidth is almost identical with that from Monte Carlo. Note that the number of our simulations (i.e., $500$) is much smaller than the total number of basis functions (i.e., $903$) in the generalized polynomial-chaos expansion.

\section{Conclusions and Future Work}
This paper has presented a big-data approach for solving the challenging high-dimensional uncertainty quantification problem. Our key idea is to estimate the high-dimensional simulation data array from an extremely small subset of its samples. This idea has been described as a tensor-recovery model with low-rank and sparse constraints. Detailed numerical methods have been described to solve the resulting optimization problem. Simulation results on a CMOS ring oscillator, a MEMS RF capacitor and an integrated photonic circuit show that our algorithm can be easily applied to problems with about $40$ to $60$ random parameters. Instead of using a huge number of (e.g., about $10^{27}$) quadrature samples, our algorithm requires only several hundreds which is even much smaller than the number of basis functions. The proposed algorithm is much more efficient than sparse grid and Monte Carlo for our tested cases, whereas Monte Carlo used to be the only feasible approach to handle the underlying high-dimensional numerical integration. 

 \begin{table} [t]
\caption{Comparison of simulation cost for the photonic circuit.} 
\centering 
\begin{tabular}{c c c c} 
\hline
method & tensor product & sparse grid & proposed \\  \thickhline % inserts single horizontal line 
simulation samples & $ 3.6\times 10^{19}$ & $3445$ & $500$ \\ \hline %inserts single line 
\end{tabular} 
\label{table:optics_cost}% is used to refer this table in the text  
\end{table}

There exist some open theoretical questions that are worth further investigations:
\begin{itemize}
	\item Firstly, it is desirable to develop a rigorous framework such that the tensor rank $r$ and the regularization parameter $\lambda$ can be determined in an optimal manner;
	\item Secondly, the resulting tensor recovery model is non-convex. A framework that can obtain its global optimal or relax the model to a convex one will be valuable;
	\item Thirdly, it is worth improving our method such that it can efficiently handle a vector output $\mathbf{\out}(\vecpar)$; 
	\item Lastly, our framework generates the subset $\Omega$ in a random way. How to generate $\Omega$ optimally is still unclear.
\end{itemize}
It is also possible to extend our framework to other engineering applications, such as power systems and robotics.

%\section*{Acknowledgment}
%This work was supported by the NSF NEEDS program and by AIM Photonics under Project MCE$\_$EPDA004. %``DFM Methods and Tools for Photonic Systems". 

\appendices
\section{Constructing Orthonormal Polynomials~\cite{Walter:1982}}
\label{subsec:uni_gPC}
Consider a single random parameter $\xi_k \in \mathbb{R}$ with a probability density function $\rho_k(\xi_k)$, one can construct a set of polynomial functions subject to the orthonormal condition:
\begin{equation}
\label{uni_gPC}
\int\limits_{\mathbb{R}}  {\phi_{k,\alpha} ( {\xi_k } )\phi_{k,\beta} ( {\xi_k } ){\rho_k}( {\xi_k } )d\xi_k }=\delta_{\alpha,\beta} \nonumber
 %\end{array} \nonumber
\end{equation}
where  $\delta_{\alpha,\beta}$ is a Delta function, integer $\alpha$ is the highest degree of $\phi_{k,\alpha} ( {\xi_k } )$. Such polynomials can be constructed as follows~\cite{Walter:1982}. Firstly, one constructs orthogonal polynomials $\{\pi_{k, \alpha}(\xi_k) \}_{\alpha=0}^p$ with an leading coefficient 1 recursively
\begin{equation}
\label{recurrence}
 \pi _{k,\alpha + 1} (\xi_k) = \left( {\xi_k - \gamma _{\alpha} } \right)\pi_{k,\alpha} (\xi_k) - \kappa _{\alpha} \pi _{k, \alpha - 1} (\xi_k)\nonumber
\end{equation}
 for $\alpha = 0,1, \ldots p-1$, with initial conditions $\pi_{k, - 1} (\xi_k) = 0$, $\pi_{k,0} (\xi_k) = 1$  and $\kappa_0=1$. For $\alpha\geq 0$, the recurrence parameters are defined as
\begin{equation}
\label{int_cal}
\begin{array}{l}
 \displaystyle{\gamma _{\alpha}  = \frac{  {\mathbb{E}\left(\xi_k\pi _{k,\alpha}^2 (\xi_k)\right)}  }{{  {\mathbb{E}\left(\pi _{k,\alpha}^2 (\xi_k)\right)} }}}, \;\displaystyle{\kappa _{\alpha+1}  = \frac{   {\mathbb{E}\left(\xi_k\pi_{k,\alpha+1}^2 (\xi_k)\right)}   }{   {\mathbb{E}\left(\xi_k\pi _{k,\alpha}^2 (\xi_k)\right)} }}. 
 \end{array}
\end{equation}
Here $\mathbb{E}$ denotes the operator that calculates expectation. Secondly, one can obtain $\{\phi_{k, \alpha}(\xi_k) \}_{\alpha=0}^p$ by normalization:
\begin{equation}
\phi_{k,\alpha}(\xi_k) = \frac{{\pi_{k,\alpha}(\xi_k)}}{{\sqrt {\kappa _0 \kappa _1  \ldots \kappa _{\alpha} } }}, \; {\rm for}\; \alpha=0,1,\ldots,p. \nonumber
\end{equation}

\section{Gauss Quadrature Rule~\cite{Golub:1969}}
\label{app:gauss_quad}
Given $\xi_k \in \mathbb{R}$ with a density function $\rho_k(\xi_k)$ and a smooth function $q(\xi_k)$,  Gauss quadrature evaluates the integral
\begin{equation}
\label{stoInt}
\int\limits_{\mathbb{R} } {q( {\xi _k } )\rho_k ( {\xi _k } )d\xi _k }  \approx \sum\limits_{i_k = 1}^{n} {q( {\xi _k^{i_k} } )} w_k^{i_k} \nonumber
\end{equation}
with an error decreasing exponentially as $n$ increases. An exact result is obtained if $q(\xi_k)$ is a polynomial function of degree $\leq 2 n-1$. One can obtain $\{(\xi_k^{i_k}, w_k^{i_k})\}_{i_k=1}^{n}$ by reusing the recurrence parameters in (\ref{int_cal}) to form a symmetric tridiagonal matrix $\mat{J} \in \mathbb{R}^{n\times n}$: 
\begin{equation}
\label{eq:jmatrix}
\mat{J}\left( {i,j} \right) = \left\{ \begin{array}{l}
 \gamma_{i - 1} ,\;{\rm{if}}\;i = j \\ 
 \sqrt {\kappa _{i} } ,\;{\rm{if}}\; i = j+ 1 \\ 
 \sqrt {\kappa _{j} } ,\;{\rm{if}}\; i = j - 1 \\ 
 0,\;{\rm{otherwise}} \\ 
 \end{array} \right.\;{\rm{for}}\;1 \le i,j \le n. \nonumber
\end{equation}
Let $\mat{J} = \mat{Q}\Sigma \mat{Q}^T$ be an eigenvalue decomposition and $\mat{Q}$ a unitary matrix, then $\xi_k^{i_k}=\Sigma(i_k,i_k)$ and $w_k^{i_k}=\left(\mat{Q}(1,i_k)\right)^2$.

\section{Assembling The Matrices and Vector in (\ref{eq:glasso})}
\label{app:glasso}
Consider the tensor factors $\mat{U}^{(1),l+1}$, $\ldots$, $\mat{U}^{(k-1),l+1}$, $\mat{X}$, $\mat{U}^{(k+1),l}$, $\ldots$, $\mat{U}^{(d),l}$ in (\ref{eq:alm_k}). We denote the $(i,j)$ element of $\mat{U}^{(k'),l}$ (or $\mat{X}$)  by scalar $u_{i,j}^{(k'),l}$ (or $x_{i,j}$), and its $j$-th column by vector $\mat{u}_{j}^{(k'),l}$ (or $\underline{\mat{x}}_j$) $\in \mathbb{R}^{n\times 1}$. Then, the cost function in (\ref{eq:alm_k}) is
\small
\begin{align}
&f \left( \ldots, \mat{U}^{(k-1),l+1}, \mat{X}, \mat{U}^{(k+1),l},\ldots \right) \nonumber\\
=& \frac{1}{2}\sum \limits_{\mat{i}\in \Omega} { \left(\sum \limits_{j=1}^r {x_{i_k,j} }\mu_{\mat{i},j}-\ten{Y} (\mat{i}) \right)^2  } +\lambda \sum \limits_{|\basisInd|\leq p} {\left|   \sum \limits_{j=1}^r {   \nu_{\basisInd, j} 	\langle \underline{\mat{x}}_j, \mat{w}_{\alpha _k}^{(k)} \rangle} \right|} \nonumber
\end{align} \normalsize
where the scalars $\mu_{\mat{i},j}$ and $ \nu_{\basisInd, j}  $ are computed as follows:
\begin{align}
\mu_{\mat{i},j} &= \prod \limits_{k'=1} ^{k-1}{  u_{i_{k'},j}^{(k'),l+1} }\prod \limits_{k'=k+1} ^{d}{  u_{i_{k'},j}^{(k'),l} }, \nonumber \\
 \nu_{\basisInd, j} &= \prod \limits_{k'=1}^{k-1} { \langle \mat{u}_{j}^{(k'),l+1}, \mat{w}_{\alpha _{k'}}^{(k')} \rangle}   \prod \limits_{k'=k+1}^{d} { \langle \mat{u}_{j}^{(k'),l}, \mat{w}_{\alpha _{k'}}^{(k')} \rangle}. \nonumber
\end{align}
 
  Since each row (or element) of $\mat{A}$ (or $\mat{b}$) corresponds to an index $\mat{i}=(i_1,\ldots, i_d)\in \Omega$, and each row of $\mat{F}$ corresponds to a basis function $\Psi_{\basisInd} (\boldsymbol {\xi})$, in this appendix we use $\mat{i}$ as the row index (or element index) of $\mat{A}$ (or $\mat{b}$) and $\basisInd$ as the row index of $\mat{F}$\footnote{We can order all elements of $\Omega$ in a specific way. If $\mat{i}$ is the $k$-th element in $\Omega$, then $\mat{A}(\mat{i},j)$ and $\mat{b}(\mat{i})$ denote $\mat{A}(k,j)$ and $\mat{b}(k)$, respectively.}. Now we specify the elements of $\mat{A}$, $\mat{b}$ and $\mat{F}$ of (\ref{eq:glasso}).
  \begin{itemize}
  \item For every $\mat{i} \in \Omega$, $\mat{b}(\mat{i})=\ten{Y} (\mat{i})$.
  
  \item Since $x_{i_k,j}$ is the $\left((j-1)n+ i_k\right)$-th element of $\mat{x}=\mat{vec}  (\mat{X}) \in \mathbb{R}^{nr\times 1}$, for every $\mat{i} \in \Omega $ we have
  \begin{equation}
\mat{A}(\mat{i},(j-1)n+i_k)= \left\{ \begin{array}{l}
 \mu_{\mat{i},j},\; {\rm{for}}\; j=1,\ldots,r \\ 
 0 ,\;\;\;\;{\rm otherwise}. 
 \end{array} \right.\nonumber
\end{equation}
  
  \item Since $\underline{\mat{x}}_j$ includes the elements of $\mat{x}  \in \mathbb{R}^{nr\times 1} $ ranging from index $(j-1)n+1$ to $jn$, given an index vector $\basisInd$ the corresponding row of $\mat{F}$ can be specified as 
  \begin{equation}
  \mat{F}(\basisInd,jn-n+i_k)= \nu_{\basisInd,j}\mat{v}_{k}^{\alpha_k}(i_k)=\nu_{\basisInd,j} \phi_{k,\alpha_k}(\xi_k^{i_k})w_k^{i_k}    \nonumber
  \end{equation}
 for all integers $j\in [1,r]$ and $i_k\in [1,n]$.
 
  \end{itemize}

\bibliographystyle{IEEEtran}
\bibliography{SPI} 

\begin{thebibliography}{10}
\providecommand{\url}[1]{#1}
\csname url@samestyle\endcsname
\providecommand{\newblock}{\relax}
\providecommand{\bibinfo}[2]{#2}
\providecommand{\BIBentrySTDinterwordspacing}{\spaceskip=0pt\relax}
\providecommand{\BIBentryALTinterwordstretchfactor}{4}
\providecommand{\BIBentryALTinterwordspacing}{\spaceskip=\fontdimen2\font plus
\BIBentryALTinterwordstretchfactor\fontdimen3\font minus
  \fontdimen4\font\relax}
\providecommand{\BIBforeignlanguage}[2]{{%
\expandafter\ifx\csname l@#1\endcsname\relax
\typeout{** WARNING: IEEEtran.bst: No hyphenation pattern has been}%
\typeout{** loaded for the language `#1'. Using the pattern for}%
\typeout{** the default language instead.}%
\else
\language=\csname l@#1\endcsname
\fi
#2}}
\providecommand{\BIBdecl}{\relax}
\BIBdecl

\bibitem{zhang2016big}
Z.~Zhang, T.-W. Weng, and L.~Daniel, ``A big-data approach to handle process
  variations: Uncertainty quantification by tensor recovery,'' in \emph{Proc.
  IEEE Workshop on Signal and Power Integrity}, 2016, pp. 1--4.

\bibitem{variation2008}
D.~S. Boning, ``Variation,'' \emph{IEEE Trans. Semiconductor Manufacturing},
  vol.~21, no.~1, pp. 63--71, Feb 2008.

\bibitem{MCintro}
S.~Weinzierl, ``Introduction to {Monte Carlo} methods,'' {NIKHEF}, Theory
  Group, The Netherlands, Tech. Rep. NIKHEF-00-012, 2000.

\bibitem{SingheeR09}
A.~Singhee and R.~A. Rutenbar, ``Statistical blockade: Very fast statistical
  simulation and modeling of rare circuit events and its application to memory
  design,'' \emph{IEEE Trans. on CAD of Integrated Circuits and systems},
  vol.~28, no.~8, pp. 1176--1189, Aug. 2009.

\bibitem{sfem}
R.~Ghanem and P.~Spanos, \emph{Stochastic finite elements: a spectral
  approach}.\hskip 1em plus 0.5em minus 0.4em\relax Springer-Verlag, 1991.

\bibitem{col:2005}
D.~Xiu and J.~S. Hesthaven, ``High-order collocation methods for differential
  equations with random inputs,'' \emph{SIAM J. Sci. Comp.}, vol.~27, no.~3,
  pp. 1118--1139, Mar 2005.

\bibitem{manfredi:tcas2014}
P.~Manfredi, D.~V. Ginste, D.~D. Zutter, and F.~Canavero, ``Stochastic modeling
  of nonlinear circuits via {SPICE}-compatible spectral equivalents,''
  \emph{IEEE Trans. Circuits Syst. I: Regular Papers}, vol.~61, no.~7, pp.
  2057--2065, July 2014.

\bibitem{Stievano:2011_1}
I.~S. Stievano, P.~Manfredi, and F.~G. Canavero, ``Parameters variability
  effects on multiconductor interconnects via hermite polynomial chaos,''
  \emph{IEEE Trans. Compon., Packag., Manufacut. Tech.}, vol.~1, no.~8, pp.
  1234--1239, Aug. 2011.

\bibitem{zzhang:tcad2013}
Z.~Zhang, T.~A. El-Moselhy, I.~A.~M. Elfadel, and L.~Daniel, ``Stochastic
  testing method for transistor-level uncertainty quantification based on
  generalized polynomial chaos,'' \emph{IEEE Trans. Computer-Aided Design
  Integr. Circuits Syst.}, vol.~32, no.~10, Oct. 2013.

\bibitem{Strunz:2008}
K.~Strunz and Q.~Su, ``Stochastic formulation of {SPICE}-type electronic
  circuit simulation with polynomial chaos,'' \emph{ACM Trans. Modeling and
  Computer Simulation}, vol.~18, no.~4, pp. 15:1--15:23, Sep 2008.

\bibitem{zzhang:tcas2_2013}
Z.~Zhang, T.~A. El-Moselhy, P.~Maffezzoni, I.~A.~M. Elfadel, and L.~Daniel,
  ``Efficient uncertainty quantification for the periodic steady state of
  forced and autonomous circuits,'' \emph{IEEE Trans. Circuits Syst. II: Exp.
  Briefs}, vol.~60, no.~10, Oct. 2013.

\bibitem{Pulch:2011_1}
R.~Pulch, ``Modelling and simulation of autonomous oscillators with random
  parameters,'' \emph{Mathematics and Computers in Simulation}, vol.~81, no.~6,
  pp. 1128--1143, Feb 2011.

\bibitem{Rufuie2014}
M.~Rufuie, E.~Gad, M.~Nakhla, R.~Achar, and M.~Farhan, ``Fast variability
  analysis of general nonlinear circuits using decoupled polynomial chaos,'' in
  \emph{Workshop Signal and Power Integrity}, May 2014, pp. 1--4.

\bibitem{yucel2015me}
A.~Yucel, H.~Bagci, and E.~Michielssen, ``An {ME-PC} enhanced {HDMR} method for
  efficient statistical analysis of multiconductor transmission line
  networks,'' \emph{IEEE Trans. Comp., Packag. Manuf. Tech.}, vol.~5, no.~5,
  pp. 685--696, 2015.

\bibitem{ahadi2015sparse}
M.~Ahadi and S.~Roy, ``Sparse linear regression ({SPLINER}) approach for
  efficient multidimensional uncertainty quantification of high-speed
  circuits,'' \emph{IEEE Trans. CAD Integr. Circ. Syst.}, vol.~35, no.~10, pp.
  1640--1652, Oct. 2015.

\bibitem{ahadi2016hyperbolic}
M.~Ahadi, A.~Prasad, and S.~Roy, ``Hyperbolic polynomial chaos expansion
  ({HPCE}) and its application to statistical analysis of nonlinear circuits,''
  in \emph{IEEE Workshop on Signal and Power Integrity}, 2016, pp. 1--4.

\bibitem{manfredi2015}
P.~Manfredi, D.~V. Ginste, D.~De~Zutter, and F.~G. Canavero, ``Generalized
  decoupled polynomial chaos for nonlinear circuits with many random
  parameters,'' \emph{IEEE Microwave and Wireless Components Letters}, vol.~25,
  no.~8, pp. 505--507, 2015.

\bibitem{zzhang_cicc2014}
Z.~Zhang, X.~Yang, G.~Marucci, P.~Maffezzoni, I.~M. Elfadel, G.~Karniadakis,
  and L.~Daniel, ``Stochastic testing simulator for integrated circuits and
  {MEMS}: Hierarchical and sparse techniques,'' in \emph{Proc. IEEE Custom
  Integrated Circuits Conf.}\hskip 1em plus 0.5em minus 0.4em\relax San Jose,
  CA, Sept. 2014, pp. 1--8.

\bibitem{zzhang:huq_tcad}
Z.~Zhang, I.~Osledets, X.~Yang, G.~E. Karniadakis, and L.~Daniel, ``Enabling
  high-dimensional hierarchical uncertainty quantification by {ANOVA} and
  tensor-train decomposition,'' \emph{IEEE Trans. CAD of Integrated Circuits
  and Systems}, vol.~34, no.~1, pp. 63 -- 76, Jan 2015.

\bibitem{twweng:optsEx}
T.-W. Weng, Z.~Zhang, Z.~Su, Y.~Marzouk, A.~Melloni, and L.~Daniel,
  ``Uncertainty quantification of silicon photonic devices with correlated and
  non-{Gaussian} random parameters,'' \emph{Optics Express}, vol.~23, no.~4,
  pp. 4242 -- 4254, Feb 2015.

\bibitem{zubac2015efficient}
Z.~Zubac, J.~Fostier, D.~De~Zutter, and D.~V. Ginste, ``Efficient uncertainty
  quantification of large two-dimensional optical systems with a parallelized
  stochastic {Galerkin} method,'' \emph{Optics Express}, vol.~23, no.~24, pp.
  30\,833--30\,850, 2015.

\bibitem{gPC2002}
D.~Xiu and G.~E. Karniadakis, ``The {Wiener-Askey} polynomial chaos for
  stochastic differential equations,'' \emph{SIAM J. Sci. Comp.}, vol.~24,
  no.~2, pp. 619--644, Feb 2002.

\bibitem{anchor_ANOVA_xiu:2012}
X.~Yang, M.~Choi, G.~Lin, and G.~E. Karniadakis, ``Adaptive {ANOVA}
  decomposition of stochastic incompressible and compressible flows,'' \emph{J.
  Comp. Phys.}, vol. 231, no.~4, pp. 1587--1614, Feb 2012.

\bibitem{anchor_ANOVA_xma:2010}
X.~Ma and N.~Zabaras, ``An adaptive high-dimensional stochastic model
  representation technique for the solution of stochastic partial differential
  equations,'' \emph{J. Comp. Phys.}, vol. 229, no.~10, pp. 3884--3915, May
  2010.

\bibitem{HDMR:1999}
H.~Rabitz and O.~Alis, ``General foundations of high-dimensional model
  representations,'' \emph{J. Math. Chem.}, vol.~25, no. 2-3, pp. 197--233,
  1999.

\bibitem{xli2010}
X.~Li, ``Finding deterministic solution from underdetermined equation:
  large-scale performance modeling of analog/{RF} circuits,'' \emph{IEEE Trans.
  CAD of Integr. Circuits Syst.}, vol.~29, no.~11, pp. 1661--1668, Nov 2011.

\bibitem{Donoho:2006}
D.~L. Donoho, ``Compressed sensing,'' \emph{IEEE Trans. Informa. Theory},
  vol.~52, no.~4, pp. 578 --594, April 2006.

\bibitem{Tarek_DAC:10}
T.~Moselhy and L.~Daniel, ``Stochastic dominant singular vectors method for
  variation-aware extraction,'' in \emph{Proc. Design Auto. Conf.}, Jun. 2010,
  pp. 667--672.

\bibitem{MoselhyD10}
T.~A. El{-}Moselhy and L.~Daniel, ``Variation-aware interconnect extraction
  using statistical moment preserving model order reduction,'' in \emph{Design,
  Automation and Test in Europe}, 2010, pp. 453--458.

\bibitem{tensor:suvey}
T.~G. Kolda and B.~W. Bader, ``Tensor decompositions and applications,''
  \emph{SIAM Review}, vol.~51, no.~3, pp. 455--500, Aug. 2009.

\bibitem{Walter:1982}
W.~Gautschi, ``On generating orthogonal polynomials,'' \emph{SIAM J. Sci. Stat.
  Comput.}, vol.~3, no.~3, pp. 289--317, Sept. 1982.

\bibitem{Gerstner:1998}
T.~Gerstner and M.~Griebel, ``Numerical integration using sparse grids,''
  \emph{Numer. Algor.}, vol.~18, pp. 209--232, Mar. 1998.

\bibitem{Davis:07}
P.~J. Davis and P.~Rabinowitz, \emph{Methods of numerical integration}.\hskip
  1em plus 0.5em minus 0.4em\relax Courier Corporation, 2007.

\bibitem{Golub:1969}
G.~H. Golub and J.~H. Welsch, ``Calculation of gauss quadrature rules,''
  \emph{Math. Comp.}, vol.~23, pp. 221--230, 1969.

\bibitem{candecomp}
J.~D. Carroll and J.~J. Chang, ``Analysis of individual differences in
  multidimensional scaling via an n-way generalization of ``{E}ckart-{Y}oung''
  decomposition,'' \emph{Psychometrika}, vol.~35, no.~3, pp. 283--319, 1970.

\bibitem{tuckerreview}
L.~R. Tucker, ``Some mathematical notes on three-mode factor analysis,''
  \emph{Psychometrika}, vol.~31, no.~3, pp. 279--311, 1966.

\bibitem{ivanTT}
I.~Oseledets, ``Tensor-train decomposition,'' \emph{SIAM J. Sci. Comp.},
  vol.~33, no.~5, pp. 2295--2317, 2011.

\bibitem{zhang:2016keynote}
Z.~Zhang, L.~Daniel, K.~Batselier, H.~Liu, and N.~Wong, ``Tensor computation: A
  new framework for high-dimensional problems in {EDA},'' \emph{IEEE Trans. CAD
  of Integr. Circuits Syst.}, {submitted in} 2016.

\bibitem{Boyd:ADMM2010}
S.~Boyd, N.~Parikh, E.~Chu, B.~Peleato, and J.~Eckstein, ``Distributed
  optimization and statistical learning via the alternating direction method of
  multipliers,'' \emph{Foundations and Trends in Machine Learning}, vol.~3,
  no.~1, pp. 1 --122, 2010.

\bibitem{zhang2014maximum}
Z.~Zhang, N.~Farnoosh, T.~Klemas, and L.~Daniel, ``Maximum-entropy density
  estimation for { MRI} stochastic surrogate models,'' \emph{IEEE Antennas and
  Wireless Propagation Letters}, vol.~13, pp. 1656--1659, 2014.

\bibitem{zzhang:JMEMS2014}
Z.~Zhang, M.~Kamon, and L.~Daniel, ``Continuation-based pull-in and lift-off
  simulation algorithms for microelectromechanical devices,'' \emph{J.
  Microelectromech. Syst.}, vol.~23, no.~5, pp. 1084--1093, Oct. 2014.

\end{thebibliography}

\begin{IEEEbiography}[{\includegraphics[width=1in,height=1.25in,clip,keepaspectratio]{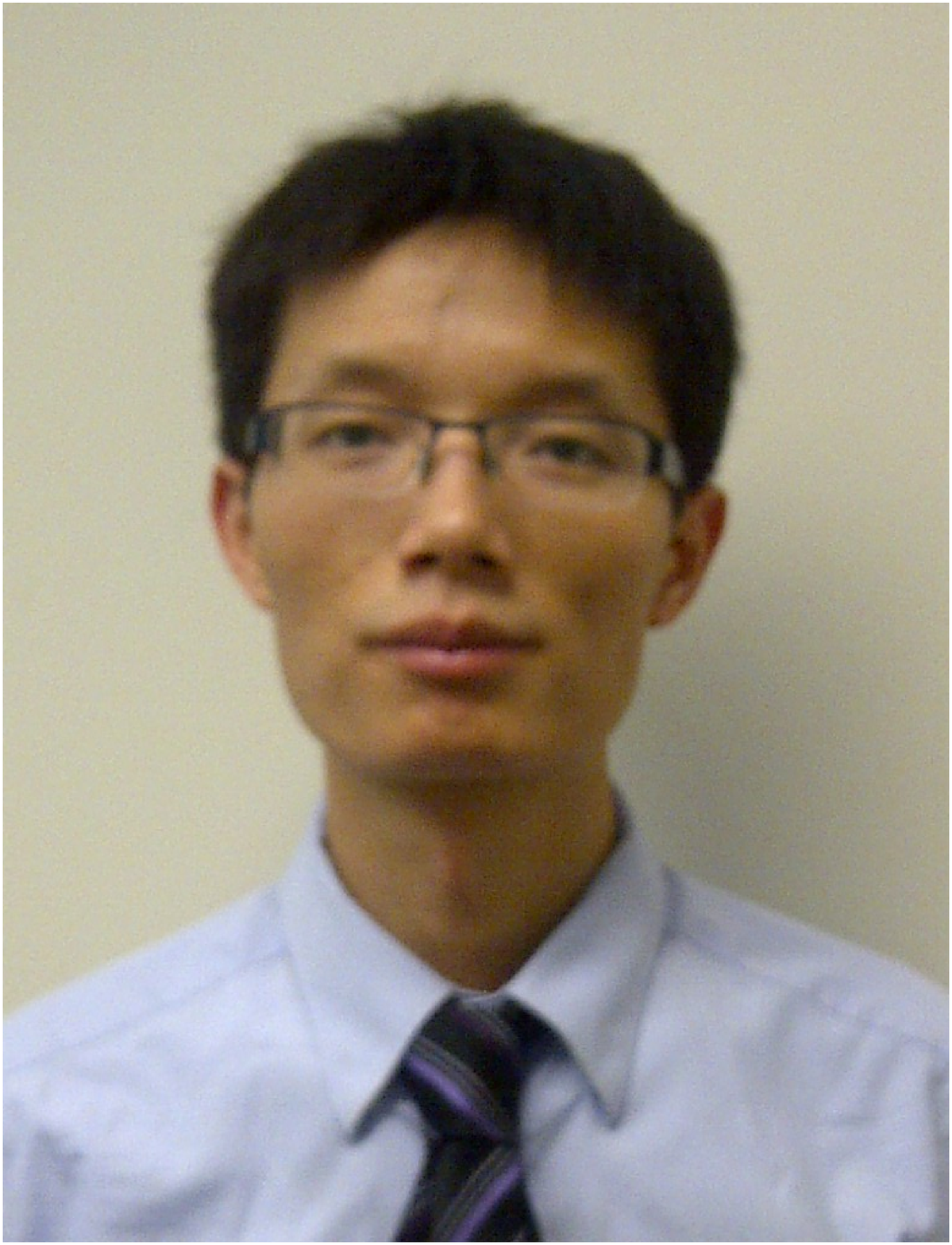}}]{Zheng Zhang} (M'15) received the Ph.D degree in Electrical Engineering and Computer Science from the Massachusetts Institute of Technology (MIT), Cambridge, MA, in 2015. Currently he is a postdoc associate with the Research Laboratory of Electronics at MIT. His research interests include uncertainty quantification, tensor and model order reduction, with application to nanoelectronics, energy and biomedical problems. His industrial experiences include Coventor Inc. and Maxim-IC; academic visiting experiences include UC San Diego, Brown University and Politechnico di Milano; government lab experiences include Argonne National Labs.

Dr. Zhang received the 2016 ACM Outstanding Ph.D Dissertation Award in Electronic Design Automation, the 2015 Doctoral Dissertation Seminar Award (i.e., Best Thesis Award) from the Microsystems Technology Laboratory of MIT, the 2014 Best Paper Award from IEEE Transactions on Computer-Aided Design of Integrated Circuits and Systems, the 2014 Chinese Government Award for Outstanding Students Abroad, and the 2011 Li Ka-Shing Prize from the University of Hong Kong. %Some of his research results have been implemented in the commerical MEMS/IC co-design software MEMS+.
\end{IEEEbiography}

\begin{IEEEbiography}[{\includegraphics[width=1in,height=1.25in,clip,keepaspectratio]{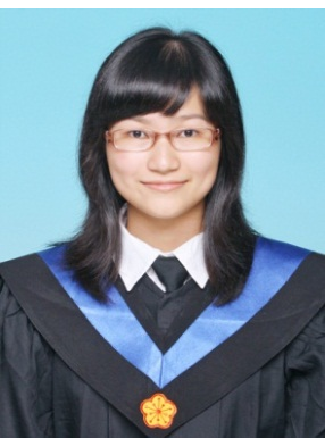}}]{Tsui-Wei Weng} (S'12) received both B.S and M.S. degrees in electrical engineering from National Taiwan University, Taipei, Taiwan, in 2011 and 2013. She is currently working toward the Ph.D degree in Department of Electrical Engineering and Computer Science at Massachusetts Institute of Technology, Cambridge, MA, USA. Her current research interests include mixed integer programming and non-convex optimization problems in machine learning, as well as uncertainty quantification in emerging technology such as artificial intelligence and nanophotonics. 
\end{IEEEbiography}

\begin{IEEEbiography}[{\includegraphics[width=1in,height=1.25in,clip,keepaspectratio]{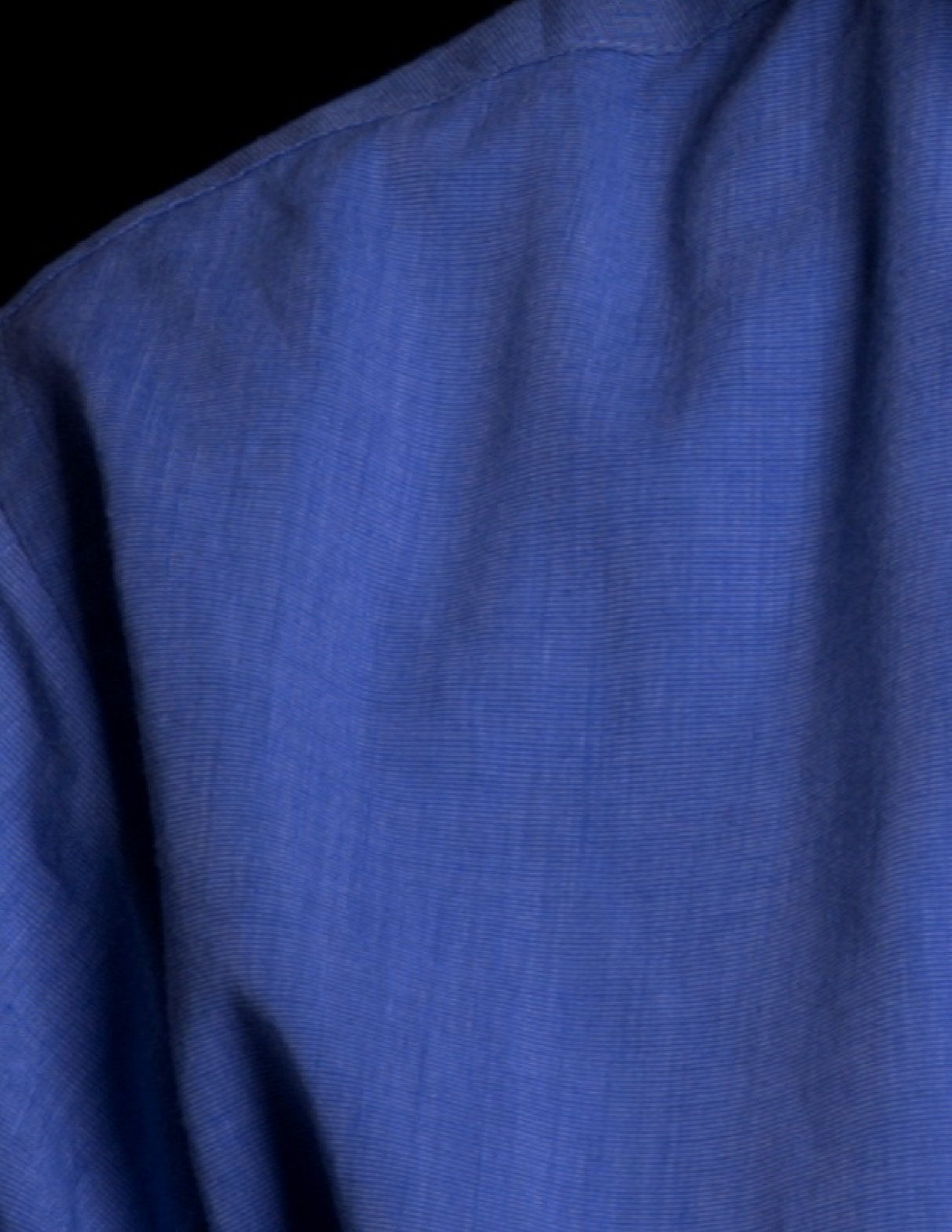}}]{Luca Daniel} (S'98-M'03) is a Full Professor in the Electrical Engineering and Computer Science Department of the Massachusetts Institute of Technology (MIT). He received the Ph.D. degree in Electrical Engineering from the University of California, Berkeley, in 2003. Industry experiences include HP Research Labs, Palo Alto (1998) and Cadence Berkeley Labs (2001).

Dr. Daniel current research interests include integral equation solvers, uncertainty quantification and parameterized model order reduction, applied to RF circuits, silicon photonics, MEMs, Magnetic Resonance Imaging scanners, and the human cardiovascular system.

Prof. Daniel was the recipient of the 1999 IEEE Trans. on Power Electronics best paper award; the 2003 best PhD thesis awards from the Electrical Engineering and the Applied Math departments at UC Berkeley; the 2003 ACM Outstanding Ph.D. Dissertation Award in Electronic Design Automation; the 2009 IBM Corporation Faculty Award; the 2010 IEEE Early Career Award in Electronic Design Automation; the 2014 IEEE Trans. On Computer Aided Design best paper award; and seven best paper awards in conferences.
\end{IEEEbiography}

\end{document}